\documentclass[twocolumn,showpacs,preprintnumbers,amsmath,amssymb]{revtex4-1}
\usepackage{amssymb}    %math formula
\usepackage{amsmath}
\usepackage{graphicx}
\usepackage[normalem]{ulem}
\usepackage{multirow}
\usepackage{appendix}
\usepackage{CJK}
\usepackage[usenames]{color}
\usepackage{bm}
\usepackage[colorlinks,linkcolor=blue,urlcolor=blue,citecolor=blue]{hyperref}
\usepackage{booktabs}     % table!
\usepackage{leftidx}
\usepackage{harpoon}
\usepackage{booktabs} 	  % table!
\usepackage{leftidx}
\usepackage{soul,xcolor}
\setstcolor{red}

\begin{document}
\begin{CJK*} {UTF8} {gbsn}

\title{System evolution of forward-backward multiplicity correlations in a multi-phase transport model }

% my name
\author{Yi-An Li(李逸安)}
\affiliation{Shanghai Institute of Applied Physics, Chinese Academy of Sciences, Shanghai 201800, China}
\affiliation{Key Laboratory of Nuclear Physics and Ion-beam Application (MOE), Institute of Modern Physics, Fudan University, Shanghai 200433, China}
\affiliation{University of Chinese Academy of Sciences, Beijing 100049, China}

% adviser name
\author{Dong-Fang Wang(王东方)}
\affiliation{Key Laboratory of Nuclear Physics and Ion-beam Application (MOE), Institute of Modern Physics, Fudan University, Shanghai 200433, China}

\author{Song Zhang(张松)}\thanks{Email: song\_zhang@fudan.edu.cn}
\affiliation{Key Laboratory of Nuclear Physics and Ion-beam Application (MOE), Institute of Modern Physics, Fudan University, Shanghai 200433, China}

\author{Yu-Gang Ma(马余刚)}\thanks{Email:  mayugang@fudan.edu.cn}
\affiliation{Key Laboratory of Nuclear Physics and Ion-beam Application (MOE), Institute of Modern Physics, Fudan University, Shanghai 200433, China}

\date{\today}
% Part 0.0: abstract
\begin{abstract}
The initial geometry effect on forward-backward multiplicity correlations $C(N_{f},N_{b})$ is studied in relativistic collisions between  light nuclei  by using a multiphase transport model (AMPT). It is found that tetrahedron $^{16}$O + $^{16}$O gives a more uniform and symmetrical fireball which produces a more isotropic  distribution of final particles  after the expansion and evolution, and leads to a small $C(N_{f},N_{b})$. Forward-backward multiplicity correlation could be taken as a useful probe to distinguish the pattern of $\alpha$-clustered $^{16}$O in experiments by comparing the neighboring colliding nuclear systems like $^{14}$N + $^{14}$N and $^{19}$F + $^{19}$F.
 \end{abstract}
\maketitle

% Part 1.0 introduction
\section{Introduction}
\par
% establish research field!
Relativistic heavy ion collisions produce an extreme hot and dense environment and provide a venue to understand the properties of early-stage quark matter in the Universe as well as the strong interaction \cite{Shuyak,PBM,Fuk,Chen,CPL2,CPL21}.  Various probes such as hard probes, soft probes and electro-magnetic probes etc \cite{NSTSongFlow,Lao2018,Lao2018b,Tang,Wang2019,Huang2020} are proposed and measurements of heavy-ion collisions at Relativistic Heavy Ion Collider (RHIC) and the Large Hadron Collider (LHC) have put important constraints on the theoretical studies and model simulation. On the other hand, the $\alpha$-clustered structure of stable nuclei $^{12}$C or $^{16}$O has been one of the highly interesting topics in the heavy-ion community at lower energy ~\cite{VONOERTZEN200643,Rus,PhysRevLett113032506,PhysRevC95034606,Huang_2017,Ye,Guo2019,Shi}, much progress has been achieved in recent years ~\cite{PhysRevLett.112.112501,PhysRevC90064902,PhysRevC97034912,PhysRevC100064912,PhysRevC95064904,Li2020,He2020,MaL2020,Cheng2019,He21,Sum21}. 
It is generally believed that the nuclear structure effect is significant only in low energy nuclear collisions. However, it was also proposed that  this kind of nuclear structure phenomenon  can be demonstrated in relativistic heavy-ion collisions through  observables such as harmonic flows~\cite{PhysRevLett.112.112501,PhysRevC90064902}.

Two general classes of quantum chromodynamics processes combine to generate events in high energy hadron collisions. Multiple parton interactions picture the soft processes, together with the beam remnants, that are characterized by long range correlations. Short range correlation, which quickly diminishes as the pseudo-rapidity distance increases, manifests the hard processes that are arisen by the perturbatively described radiation or partonic scatterings and produced by single- and few-parton exchanges~\cite{Aad2012}.

Due to the strong initial-state density fluctuations in the light nucleus, the space-time evolution of the produced matter in the final state fluctuates event to event. These density fluctuations generate long-range correlations (LRC) at the early stages of the collision, well before the onset of any collective behavior, and appear as correlations of the multiplicity densities of produced particles separated in pseudo-rapidity ($\eta $)~\cite{PhysRevC.95.064914,BIALAS2012332,PhysRevC.87.024906,PhysRevC.90.034915}. The forward-backward (FB) correlation between final-state charged particle multiplicities in two separated $\eta $ windows is a useful observable in high-energy hadron or nuclear collisions to study the dynamics of particle production mechanism~\cite{BRAVINA2018146,PhysRevC.79.034910,UHLIG197815,PhysRevLett.103.172301} and may provide the information of collided nuclei if it is built up with exotic nuclear structure, such as $\alpha$-cluster. 

With that in mind, the forward-backward correlations are investigated through the AMPT model in different collision systems at center of mass energy $\sqrt{s_{NN}}$ = 200 GeV and 6370 GeV. We find that system scan experiment could be a good way to distinguish the exotic $\alpha$-clustered nuclear structure from the Woods-Saxon one.

The rest of the paper is arranged as follows: In section~\ref{sec:model} and section~\ref{sec:FB} a brief introduction to AMPT model and definition of forward-backward correlation is presented, respectively. Section \ref{sec:results and discussion} presents the centrality,  system  as well as sphericity dependences of FB multiplicity correlation. 
%Sec.~\ref{sec:sphe} shows  dependence of FB multiplicity correlation.  Sec.~\ref{sec:alph} 
More importantly, the section describes how to distinguish the structure of light nuclei through FB multiplicity correlation, %$b_\text{corr}$,
and then a summary  is given in Section ~\ref{sec:summary}.

\section{a brief introduction to AMPT model}
\label{sec:model}
A multi-phase transport model is developed to address the non-equilibrium many-body dynamics and aims at describing physics in relativistic heavy-ion collisions at RHIC~\cite{PhysRevC.72.064901,Lin2021}. It is also suitable to reproduce the results at LHC~\cite{AMPTGLM2016} including the pion HBT correlations~\cite{AMPTHBT}, dihadron azimuthal correlations~\cite{AMPTDiH}, collective flow~\cite{STARFlowAMPT,AMPTFlowLHC,PhysRevC101021901}, and strangeness production and correlation~\cite{SciChinaJinS,WangDF}. AMPT is a hybrid dynamic transport model, which consists of four main components: (a) the initial conditions including the spatial and momentum distributions of mini-jet partons and soft string excitation, which are obtained from the HIJING model; (b) partonic cascade~\cite{ZPCModel}, whereby interactions among partons are described by equations of motion for their Wigner distribution functions; (c) hadronization, which is conversion from the partonic to the hadronic matter; and (d) hadronic interactions, based on the ART (a relativistic transport) model~\cite{ARTModel}, including baryon-baryon, baryon-meson, and meson-meson elastic and inelastic scatterings. Details of the AMPT model can be found in recent reviews~\cite{PhysRevC.72.064901,Lin2021}.

The initial nucleon distribution in nuclei is configured in the HIJING model~\cite{HIJING-1,HIJING-2} with either a pattern of Woods-Saxon distribution or an exotic nucleon distribution which is embedded to study the $\alpha$-clustered structure of $^{16}$O. For details, a tetrahedral four-$\alpha$ clustering structure of $^{16}$O is taken into account in this work and  parameters of the tetrahedral structure %of $^{16}\mathrm{O}$ 
are inherited from an extended quantum molecular dynamics (EQMD) model~\cite{PhysRevLett113032506}, which is extended from the quantum molecular dynamics (QMD) model. With the effective Pauli potential, EQMD model can give reasonable $\alpha$-cluster configurations for 4$N$ nuclei. For the four $\alpha$s in the tetrahedral structure, we put them at the vertices with side length of 3.42 fm so that it gives a similar rms-radius (2.699 fm) to the Woods-Saxon configuration (2.726 fm) as well as the experimental data (2.6991 fm)~\cite{ORMSRExp}, while nucleons inside each $\alpha$ are initialized by using the Woods-Saxon distribution introduced in the HIJING model. 

\section{forward-backward multiplicity correlation}
\label{sec:FB}

%\subsection{Definition and Notations}
%\label{sec:Definition and Notations}

One method to characterize the forward-backward multiplicity correlation can be described by the following definition \cite{PhysRevC.66.044904,PhysRevC.101.014907}
\begin{equation}
\begin{aligned}
C(N_{f},N_{b}) = \frac{\left \langle N_{f}N_{b} \right \rangle-\left \langle N_{f} \right \rangle\left \langle N_{b} \right \rangle}{\left \langle N_{f} \right \rangle\left \langle N_{b} \right \rangle}
\label{equation1}
\end{aligned},
\end{equation}
 where $N_{f}$ and $N_{b}$ are the numbers of charged particles falling into the forward and backward pseudorapidity interval $\delta \eta $, respectively.  
The quantity $C(N_{f},N_{b})$ vanishes if there is no correlation between $N_{f}$ and $N_{b}$, so that $C(N_{f},N_{b})$ measures the deviation from Poisson-statistical behavior. Of practical importance, the ratio is robust since it is independent of experimental efficiency as well as what fraction of particles are used.

Here two intervals separated symmetrically around $\eta = 0$ with pseudorapidity width $\delta \eta $ are defined as ``forward" ($\eta >0$) and ``backward" ($\eta <0$). 
Correlations between multiplicities of charged particles are studied as a function of the gap between the windows $\eta_\text{gap}$, i.e. the distance between lower and upper boundary of forward and backward $\eta $ window.

Reference multiplicities are used here to reduce the influence of centrality selection on forward-backward multiplicity correlations. The parameters are set as $\delta \eta = 0.2$, and $\eta_\text{gap}$ = 0, 0.2, 0.4, 0.6, 0.8, 1.0, 1.2, 1.4 and 1.6. For $\eta_\text{gap}$ = 0, 0.2 and 0.4, reference multiplicity is set in $0.5<\left | \eta  \right |<1.0$. For $\eta _\text{gap}$ = 0.6 and 0.8, the reference multiplicity is the sum of multiplicities in $\left | \eta  \right |<0.3$ and $0.8<\left | \eta  \right |<1.0$. While for $\eta_\text{gap}$ = 1.0, 1.2, 1.4 and 1.6, reference multiplicity is obtained from $\left | \eta  \right |<0.5$. The similar approach can be found in Refs.~\cite{PhysRevLett.103.172301,PhysRevC.93.044918,PhysRevC.88.044903}.

\section{Results and discussion}
\label{sec:results and discussion}

\subsection{System and Centrality dependence of FB correlation}

\begin{figure*}[htb]
	\includegraphics[angle=0,scale=0.9]{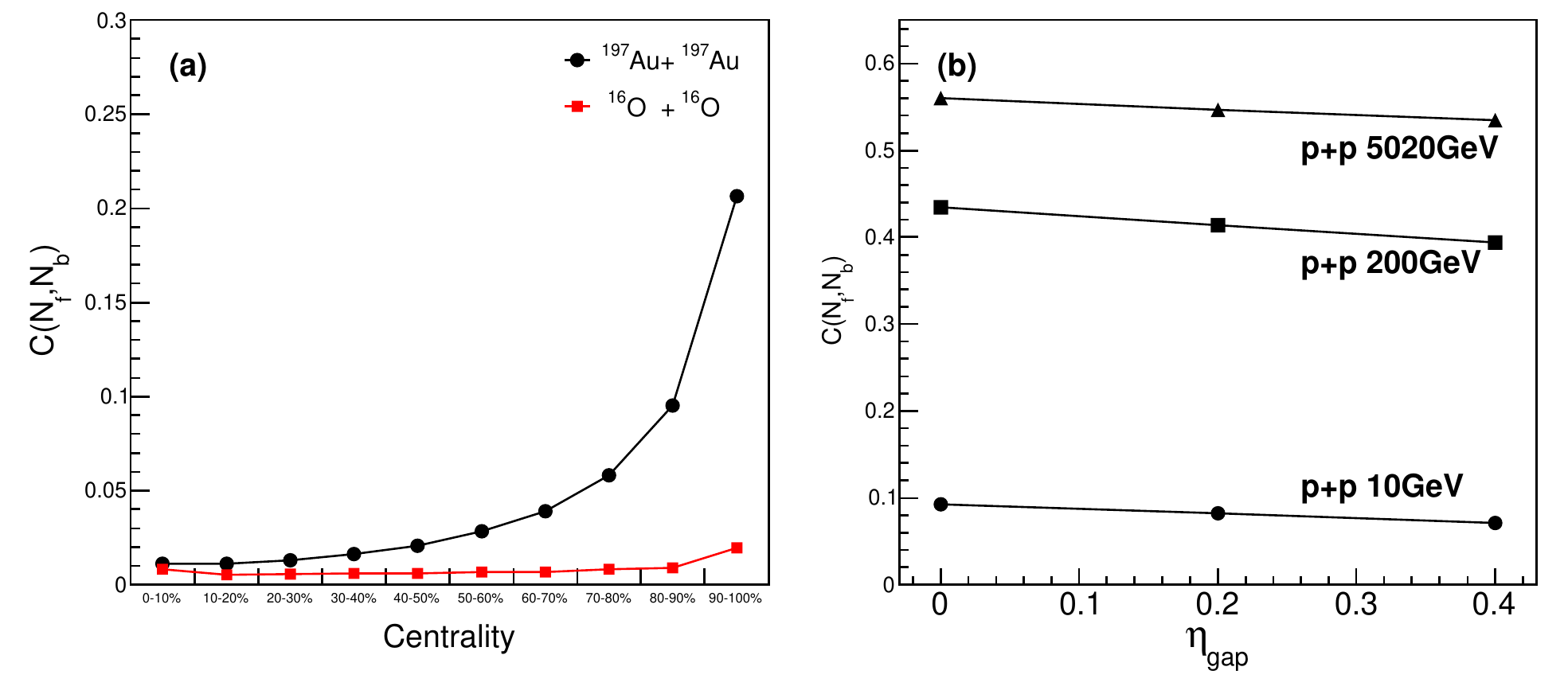}
	\caption{(a) FB multiplicity correlation coefficient $C(N_{f},N_{b})$ as a function of centrality in $^{197}$Au + $^{197}$Au collisions at $\sqrt{s_{NN}}$ = 200 GeV as well as $^{16}$O + $^{16}$O collisions at $\sqrt{s_{NN}}$ = 6370 GeV, (b) FB multiplicity correlation coefficient $C(N_{f},N_{b})$ as a function of $\eta_\text{gap}$ in pp collisions at $\sqrt{s_{NN}}$ = 10, 200, and 5020 GeV.}
	\label{fig1}
\end{figure*}

\begin{table*}[]
	\scriptsize
	\centering
	\caption{Parameters of final state charged particles in different collision systems.}
	\label{AMPT_final}
	
	\begin{tabular}{c|ccc|ccc}
		\toprule
		\multicolumn{1}{c}{} & \multicolumn{3}{c}{$\mathrm{\leftidx{^{197}}Au}+\mathrm{\leftidx{^{197}}Au}$ @ $\sqrt{s_{NN}}$ = 200 GeV}  & \multicolumn{3}{c}{$\mathrm{\leftidx{^{16}}O}+\mathrm{\leftidx{^{16}}O}$ @ $\sqrt{s_{NN}}$ = 6370 GeV} \\
		centrality & average &std ($\sigma$) &$std/average$
	   & average &std ($\sigma$) &$std/average$ \\
		\hline
		 0$\%$-10$\%$ & 93.23 & 13.69 & 0.147 & 27.59 & 6.728 & 0.244 \\
	   10$\%$-20$\%$ & 67.80 & 10.79 & 0.159 & 20.22 & 5.093 & 0.252 \\
	   20$\%$-30$\%$ & 49.29 & 8.854 & 0.180 & 15.60 & 4.443 & 0.285 \\
	   30$\%$-40$\%$ & 35.21 & 7.356 & 0.209 & 12.11 & 3.873 & 0.320 \\
	   40$\%$-50$\%$ & 24.34 & 5.990 & 0.246 & 9.325 & 3.404 & 0.365 \\
	   50$\%$-60$\%$ & 16.23 & 4.809 & 0.296 & 7.024 & 2.954 & 0.421 \\
	   60$\%$-70$\%$ & 10.31 & 3.766 & 0.365 & 5.112 & 2.524 & 0.494 \\
	   70$\%$-80$\%$ & 6.231 & 2.873 & 0.461 & 3.512 & 2.100 & 0.598 \\
	   80$\%$-90$\%$ & 3.481 & 2.140 & 0.615 & 2.083 & 1.622 & 0.779 \\
	 90$\%$-100$\%$ & 1.523 & 1.449 & 0.951 & 0.940 & 1.097 & 1.167 \\
		\bottomrule
	\end{tabular}
\end{table*}

Figure~\ref{fig1} (a) shows the FB multiplicity correlation coefficient $C(N_{f},N_{b})$ as a function of centrality in $^{197}$Au + $^{197}$Au collisions at $\sqrt{s_{NN}}$ = 200 GeV and $^{16}$O + $^{16}$O collisions at $\sqrt{s_{NN}}$ = 6370 GeV. 
The centrality of the collision is characterized by %the $N_{cp}$ (
number of final state charged particles including  $\pi ^{\pm }$, $K^{\pm }$, p and $\bar{p}$. %)
It is observed that in $^{197}$Au + $^{197}$Au collisions at $\sqrt{s_{NN}}$ = 200 GeV, $C(N_{f},N_{b})$ increases with the centrality, i.e. from central to peripheral collisions, while in $^{16}$O + $^{16}$O collisions at $\sqrt{s_{NN}}$ = 6370 GeV, it remains almost unchanged.
Figure~\ref{fig1} (b) shows the FB multiplicity correlation coefficient $C(N_{f},N_{b})$ as a function of $\eta _\text{gap}$ in $p+p$ collision at three collision energies $\sqrt{s_{NN}}$ = 10, 200, and 5020 GeV. At each collision energy, $C(N_{f},N_{b})$ is found to decrease slowly with the increasing of $\eta_\text{gap}$. It is found that the value of $C(N_{f},N_{b})$ increases with the collision energy. In $p+p$ collision, the mean multiplicity in pseudorapidity window $0< \eta  <0.8$ is around 1  at $\sqrt{s_{NN}}$ = 10 GeV, and it is around 3.5  at $\sqrt{s_{NN}}$ = 5020 GeV. The quite large values of $C(N_{f},N_{b})$ in $p+p$ collision systems show a strong correlation between forward bin and backward bin, which indicates that jet events could  play an important role in $C(N_{f},N_{b})$. This argument also suggests that  we should focus on the shape of final particle distribution in each event.

Table~\ref{AMPT_final} shows the mean, the standard deviation ($std$) and their ratio of final state charged particles of Figure~\ref{fig1} (a) in pseudorapidity window $0< \eta  <0.2$. It is observed that, with the increase of centrality, the mean value of final state charged particles decreases sharply in $^{197}$Au + $^{197}$Au system at $\sqrt{s_{NN}}$ = 200 GeV and in $^{16}$O + $^{16}$O system at $\sqrt{s_{NN}}$ = 6370 GeV. It is also found that the FB multiplicity correlation coefficient $C(N_{f},N_{b})$ is closely related to relative fluctuation when the multiplicity of final state charged particles is large, which suggests that we can use the relative fluctuation to understand the $C(N_{f},N_{b})$ difference when the system size is comparable.
After classification by centrality in $^{16}$O + $^{16}$O collisions, both $N_{f}$ and $N_{b}$ are confined to a fairly small multiplicity range, which leads $C(N_{f},N_{b})$ to be very close to 0.

\begin{figure}[htb]
	\includegraphics[angle=0,scale=0.45]{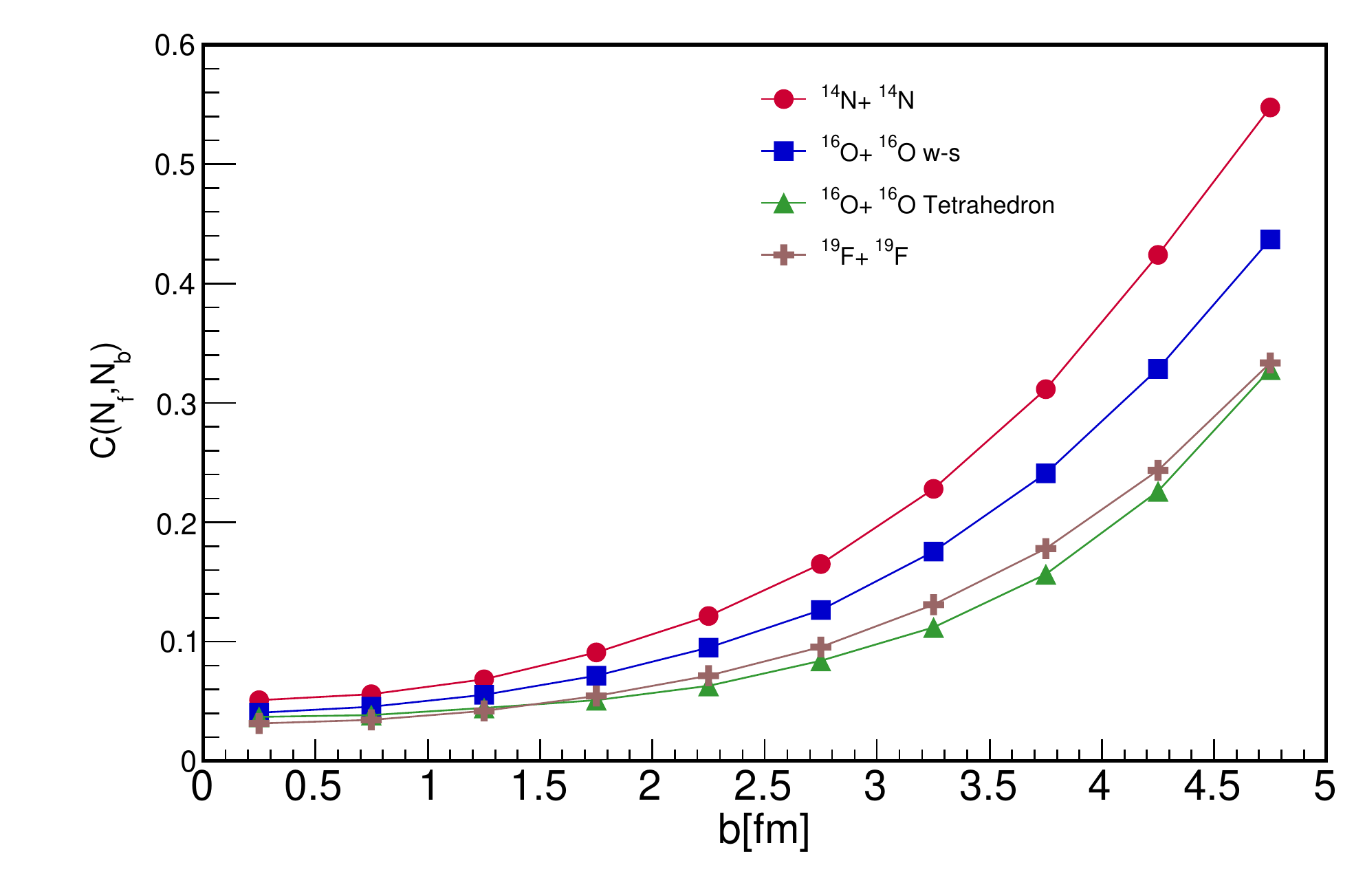}
	\caption{FB multiplicity correlation coefficient $C(N_{f},N_{b})$ as a function of impact parameter $b$ for different collision systems (a) $^{14}$N + $^{14}$N, (b) $^{16}$O + $^{16}$O, and (c) $^{16}$O + $^{16}$O with tetrahedral configuration, (d) $^{19}$F + $^{19}$F at $\sqrt{s_{NN}}$ = 6370 GeV.}
	\label{fig2}
\end{figure}

\begin{figure}[htb]
	\includegraphics[angle=0,scale=0.45]{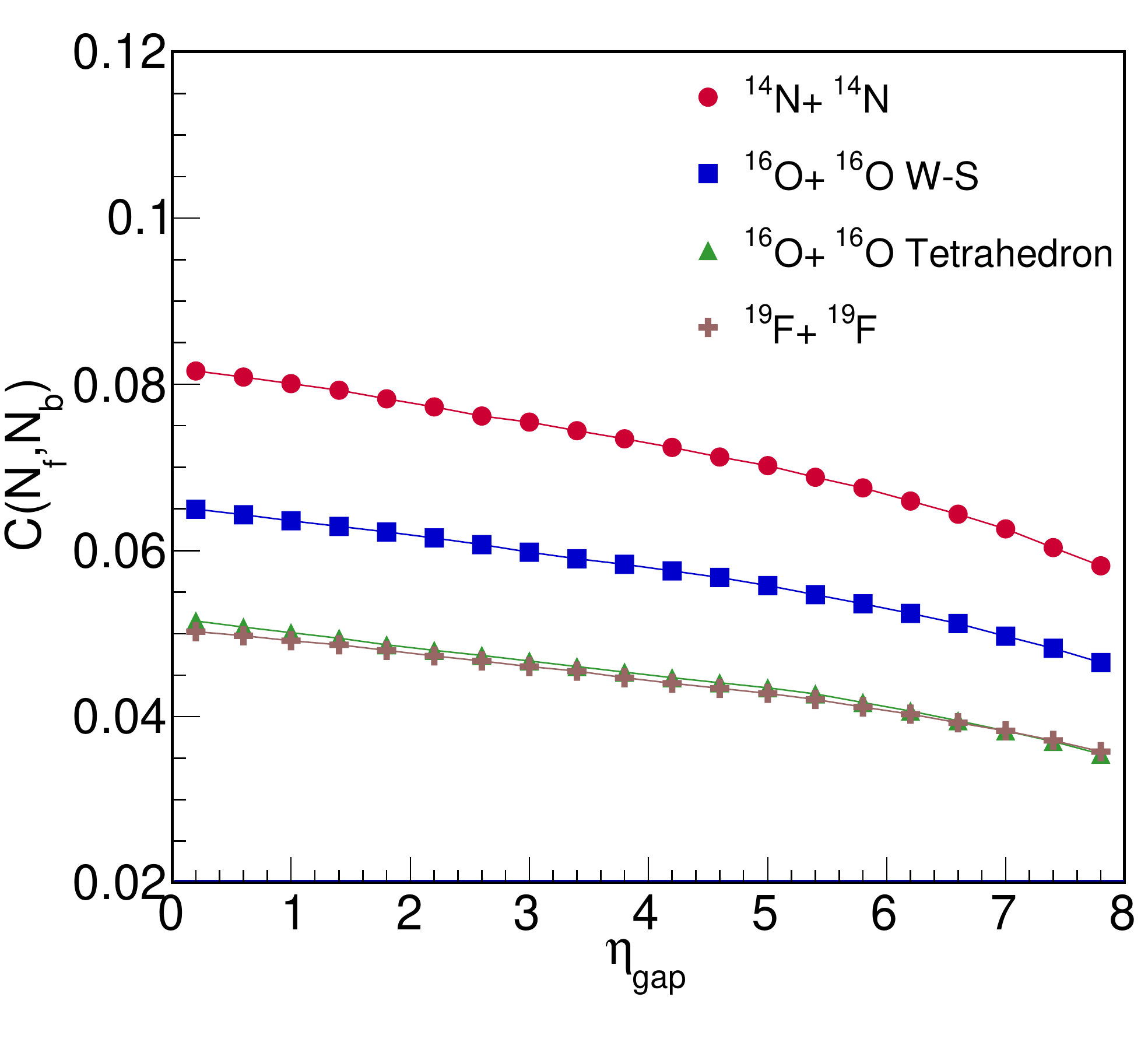}
	\caption{The correlation $C(N_{f},N_{b})$ as a function of $\eta_\text{gap}$ with window widths $\delta \eta$ = 0.2 for different collision systems at $\sqrt{s_{NN}}$ = 6370 GeV.}
	\label{fig3}
\end{figure}

To illustrate the FB correlation above as a probe to distinguish the clustering light nuclei, we also compute the forward-backward multiplicity correlation $C(N_{f},N_{b})$ of different initial structures of $^{16}$O as well as its neighbouring collisions systems, namely $^{14}$N + $^{14}$N and $^{19}$F + $^{19}$F. 
Figure~\ref{fig2} shows the FB multiplicity correlation coefficient $C(N_{f},N_{b})$ as a function of impact parameter $b$ for different collision systems at $\sqrt{s_{NN}}$ = 6370 GeV. In each collision system, $C(N_{f},N_{b})$ increases with the impact parameter $b$, which can be understood through that the relative fluctuation increases with $b$. It can be seen that the correlation functions $C(N_{f},N_{b})$ present different correlation strength in $^{16}$O+$^{16}$O collisions with $^{16}$O configuration either in Woods-Saxon distribution or $\alpha$-clustered tetrahedron structure.
To clearly investigate the effect from $\alpha$-clustering structure, figure~\ref{fig3} shows $C(N_{f},N_{b})$ as a function of $\eta_\text{gap}$ with  $b<2.0$ fm for different collision systems at $\sqrt{s_{NN}}$ = 6370 GeV. With the increasing system size, for light nuclei with Woods-Saxon distribution the $C(N_{f},N_{b})$ gradually decreases, close to zero, which can be understood that large systems tend to be binomial distribution, and the correlation between $N_{f}$ and $N_{b}$ becomes weaker.

\subsection{Sphericity dependence of FB correlation}

Considering that the shape of event has a great effect on the FB multiplicity correlation strength in $p+p$ collision, to understand the source of multiplicity correlation in A+A collisions, it is necessary to perform a sphericity dependent analysis of  $C(N_{f},N_{b})$ here. Transverse sphericity is defined as~\cite{Banfi2010}
\begin{equation}
\begin{aligned}
S_{\perp }^\text{phero}\equiv \frac{\pi ^{2}}{4}\underset{\widehat{n}=(n_{x},n_{y},0)}{\min}(\frac{\sum_{i}\left | \overrightarrow{p_{\perp ,i}}\times \widehat{n} \right |}{\sum _{i}p_{\perp ,i}})^{2}
\label{equation3}
\end{aligned},
\end{equation}
where $\overrightarrow{p_{\perp}}$ is the transverse momentum and $\widehat{n}$ is the unit vector which minimizes $S_{\perp }^{\text{phero}}$. Numerically, the minimization is simplified by the observation that the $\widehat{n}$ provides the minimal sum always coincides with the direction of one of the $\overrightarrow{p_{\perp}}$. Sphericity variable allows us to distinguish isotropic or spherical events which have high value of $S_{\perp }^{\text{phero}}$ from the pencil shaped events which have low value of $S_{\perp }^{\text{phero}}$ (eg. events are dominated with dijets).

\begin{figure}[htb]
	\includegraphics[angle=0,scale=0.45]{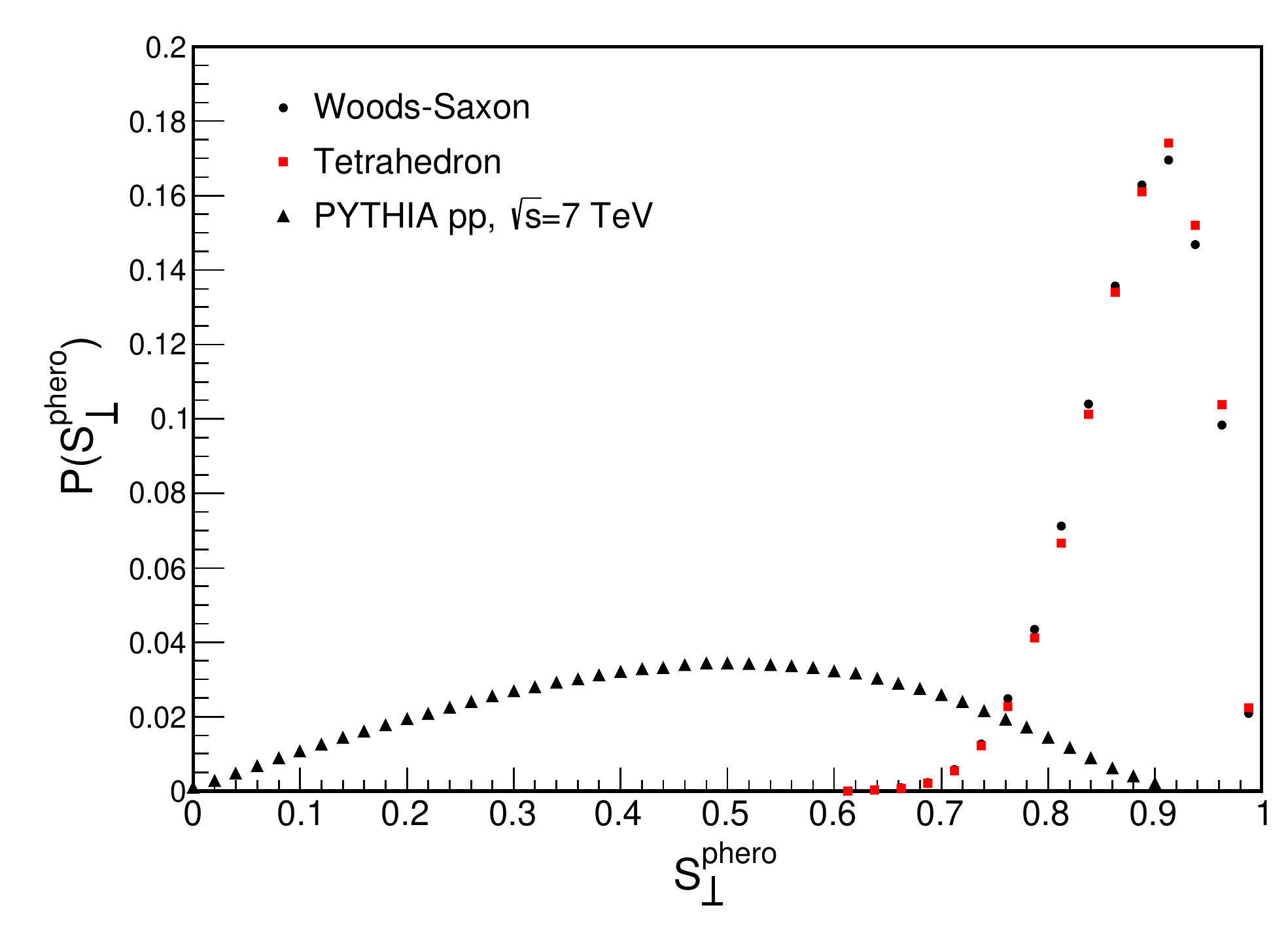}
	\caption{Sphericity probability distribution in p + p collision at $\sqrt{s_{NN}}$ = 7 TeV as well as in $^{16}$O +  $^{16}$O collisions for the most central collision at  $\sqrt{s_{NN}}$ = 6370 GeV. 
	}
	\label{fig4}
\end{figure}

\begin{figure}[htb]
	\includegraphics[angle=0,scale=0.45]{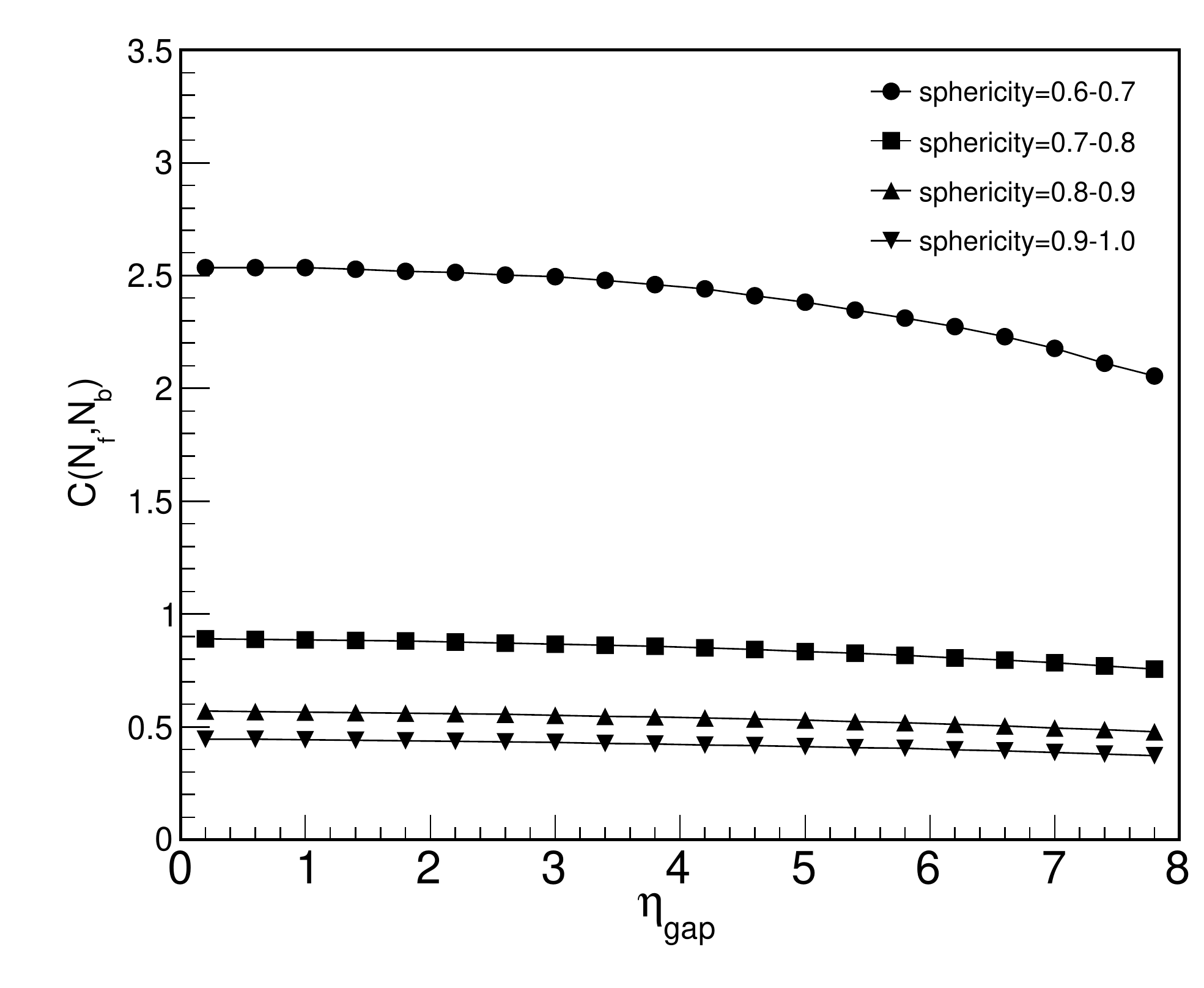}
	\caption{The correlation $C(N_{f},N_{b})$ as a function of sphericity for different collision systems at $\sqrt{s_{NN}}$ = 6370 GeV.}
	\label{fig5}
\end{figure}

Different from the sphericity distribution in p + p collisions as shown in Fig.~\ref{fig4}, which starts at $S_{\perp }^{\text{phero}} = 0$, reaches its maximum at $S_{\perp }^{\text{phero}} = 0.5$, and ends at $S_{\perp }^{\text{phero}} = 0.9$~\cite{kundu2019effect},  the sphericity distribution for charged particles in $^{16}$O +  $^{16}$O at $\sqrt{s_{NN}}$ = 6370 GeV shows a more spherical picture: there is no distribution when $S_{\perp }^{\text{phero}}<0.6$, reaches its maximum at $S_{\perp }^{\text{phero}} = 0.9$, and ends at $S_{\perp }^{\text{phero}} = 1$. From Figure~\ref{fig5}, we observed the events with $0.6<S_{\perp }^{\text{phero}}<0.7$ give the biggest correlation $C(N_{f},N_{b})$ in $^{16}$O +  $^{16}$O collisions. When the events become more isotropic like or spherical like, $C(N_{f},N_{b})$ goes down. The physical picture is that a uniform spread of final state particles gives a small $C(N_{f},N_{b})$.

%\begin{widetext}
\begin{table*}[htb]
	\scriptsize
	\centering
	\caption{Participant nucleon parameters of different collision systems in AMPT model.}
	\label{AMPT_initial}
	
	\begin{tabular}{c|ccc|cc}
		\toprule
		System & $\langle N_{part}^P\rangle$ & $\sigma(N_{part}^P)$ & $\sigma(N_{part}^P)/\langle N_{part}^P\rangle$
		& $\sigma(N_{part}^P-N_{part}^T)$ & $\sigma(N_{part}^P-N_{part}^T)/\langle N_{part}^P\rangle$ \\
		\hline
		$\mathrm{\leftidx{^{14}}N} + \mathrm{\leftidx{^{14}}N}$ & 12.11 & 1.571 & 0.130 & 1.789 & 0.148 \\
		$\mathrm{\leftidx{^{16}}O} + \mathrm{\leftidx{^{16}}O}$ & 14.10 & 1.598 & 0.113 & 1.859 & 0.132 \\
		$\mathrm{\leftidx{^{16}}O}* + \mathrm{\leftidx{^{16}}O}*$ & 14.31 & 1.395 & 0.097 & 1.697 & 0.119 \\
		$\mathrm{\leftidx{^{19}}F} + \mathrm{\leftidx{^{19}}F}$ & 16.99 & 1.659 & 0.098 & 1.969 & 0.116 \\
		\bottomrule
	\end{tabular}
\end{table*}
%\end{widetext}

Table~\ref{AMPT_initial} shows the statistics of the number of participants corresponding to the systems in figure~\ref{fig3}, $N_{part}^P$ and $N_{part}^T$ denotes the number of participants in projectile and target, respectively, $\sigma(X)$ means the deviation of $X$, $\langle\cdots\rangle$ represents the mean value. Since we are dealing with symmetric systems, only  projectile side is shown  for simplicity. 

As we can see from the Table~\ref{AMPT_initial}, with increasing of system size, $i.e.$ the Woods-Saxon configured systems from $^{14}$N + $^{14}$N, $^{16}$O + $^{16}$O, to $^{19}$F + $^{19}$F, the relative fluctuation of participant nucleon number becomes smaller. Fireball after collision is more uniform and symmetrical for the tetrahedral configuration $^{16}$O in comparison with the Woods-Saxon one.

Assuming the distribution is binomial, its mean value should be proportional to system size $O(N)$, and its fluctuation is proportional to the $\sqrt{O(N)}$, and then the relative fluctuation is proportional to $1/\sqrt{O(N)}$. In this context, for the  Woods-Saxon  configured systems from $^{14}$N + $^{14}$N, $^{16}$O + $^{16}$O, to $^{19}$F + $^{19}$F, the relative fluctuation becomes smaller with the increasing of system size. Observing  that the particle number in a grand canonical ensemble in thermal equilibrium follows Poisson statistics, so $C(N_{f},N_{b})$ becomes smaller with the system size. Comparing the systems of tetrahedron and  Woods-Saxon  $^{16}$O + $^{16}$O, the mean values are similar but the distribution width (so the standard deviation of the number of final state charged particles) of tetrahedron configuration is narrower, which could explain the correlation coefficient $C(N_{f},N_{b})$ of $^{16}\mathrm{O}$ with tetrahedral structure is slightly smaller than that with Woods-Saxon structure. 

\begin{figure*}[htb]
	\includegraphics[angle=0,scale=0.85]{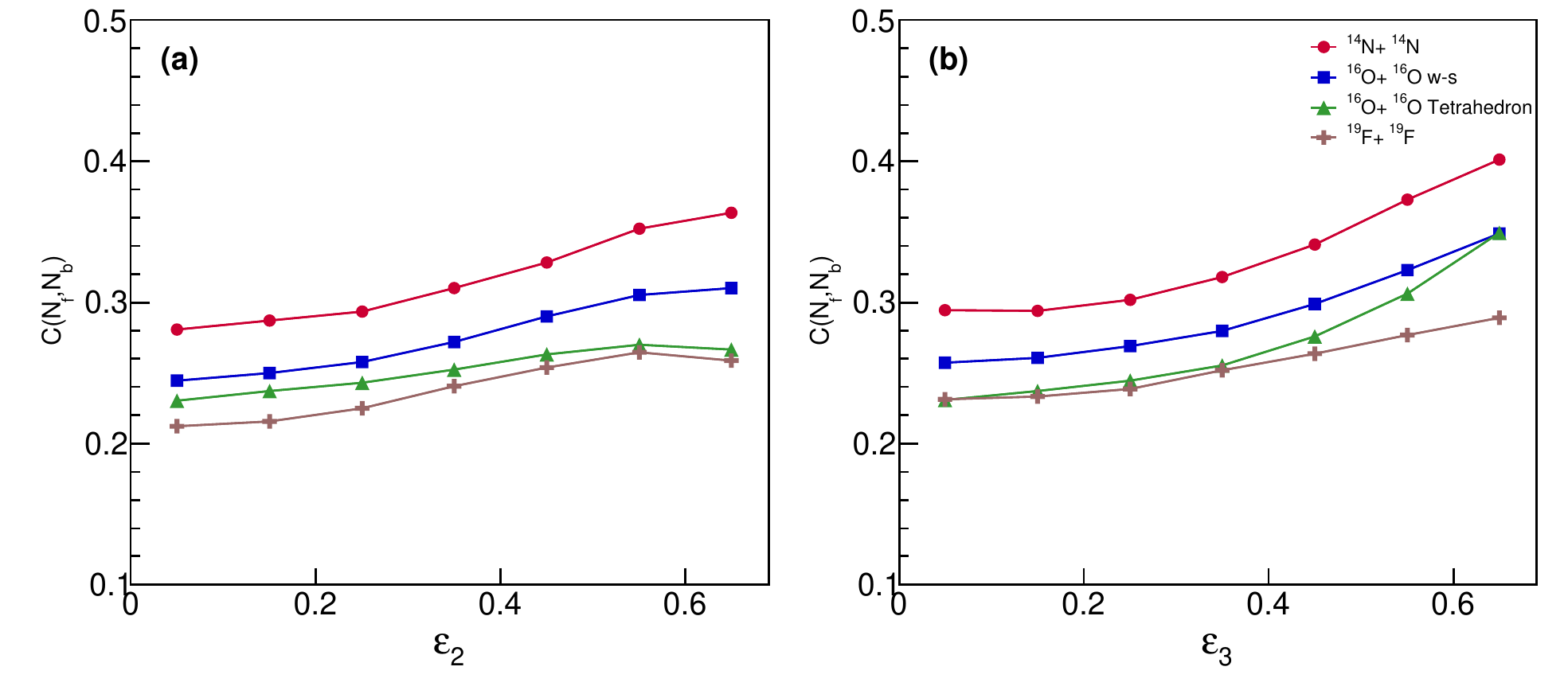}
	\caption{The correlation $C(N_{f},N_{b})$ as a function of $\varepsilon_2$ and $\varepsilon_3$ for different collision systems at $\sqrt{s_{NN}}$ = 6370 GeV.}
	\label{fig6}
\end{figure*}

\subsection{ Eccentricity dependence of FB correlation}

Considering that elliptic anisotropy of the collective component in both rapidity bins is highly correlated~\cite{PhysRevLett.103.042302} and we are now discussing the relationship between initial configuration and $C(N_{f},N_{b})$, it is natural to check the initial anisotropic density characterized by complex eccentricity coefficients ($\varepsilon_n$) in the transverse plane.
Figure~\ref{fig6} shows $C(N_{f},N_{b})$ as a function of $\varepsilon_2$ and $\varepsilon_3$ for different collision systems at $\sqrt{s_{NN}}$ = 6370 GeV. $\varepsilon_{n}e^{in\Phi_{n}} = -\left \langle r^{n}e^{in\phi } \right \rangle/\left \langle r^{n} \right \rangle$, where $r = \sqrt{x^{2}+y^{2}}$ and $\phi$ are the coordinate position and azimuthal angle of initial-state participant nucleons.
$\Phi_{n}$ is the initial participant plane angle. 
As we expected, from top to bottom are the $^{14}$N + $^{14}$N, $^{16}$O + $^{16}$O, and $^{19}$F + $^{19}$F systems with Woods-Saxon configuration. The larger system is easier to achieve thermal equilibrium, which gives a smaller $C(N_{f},N_{b})$. Focus on one system, the proportion of peripheral collision events in each bin is increasing from left to right, as we had expected more jet-like events in the final state gives larger $C(N_{f},N_{b})$.

From Table~\ref{AMPT_initial} we found that comparing with Woods-Saxon configuration, tetrahedron $^{16}$O + $^{16}$O gives a more uniform and symmetrical fireball, after the expansion and evolution it shows a more isotropic or spherical final particle distribution just as Figure~\ref{fig4} reveals. Events with spherical distribution give a small $C(N_{f},N_{b})$, whether in $p+p$ collision or in $^{16}$O + $^{16}$O and $^{197}$Au + $^{197}$Au system. It seems that $C(N_{f},N_{b})$ is quite sensitive to the distribution shape of final state particles. We also showed adjacent nuclei $^{19}$F + $^{19}$F that could be used in experiments, the $C(N_{f},N_{b})$ of tetrahedron $^{16}$O + $^{16}$O is hopeful similar to or even smaller than that of $^{19}$F + $^{19}$F, which makes it possible for us to distinguish the ground state structure of $^{16}$O experimentally.

Based on the behavior of $C(N_{f},N_{b})$, we argue that the forward-backward multiplicity correlation measurement shall be a feasible observable to distinguish the $\alpha$-clustering structure through the systematic measurement for $^{16}$O +  $^{16}$O collision and its neighbouring nuclear collisions. 

\section{summary}
\label{sec:summary}
In summary, a systematic study on forward-backward multiplicity correlations $C(N_{f},N_{b})$ from large systems to small ones has been performed through the AMPT model. It is observed that the magnitude of FB correlation strength decreases from central to peripheral in Au + Au collisions at $\sqrt{s_{NN}}$ = 200 GeV
and the $C(N_{f},N_{b})$ increases with the energy from $\sqrt{s_{NN}}$ = 10 GeV, 200 GeV to 5.02 TeV is observed in $p+p$ collision.

Considering the shape of event has a great effect on the FB multiplicity correlation strength in $p+p$ collision, a sphericity dependent study of $C(N_{f},N_{b})$ is performed in $^{16}$O +  $^{16}$O system and  $C(N_{f},N_{b})$ is found highly dependent on $S_{\perp }^{\text{phero}}$. When the events become more isotropic like, $C(N_{f},N_{b})$ becomes smaller, that's to say a uniform spread of final state particles gives a small $C(N_{f},N_{b})$.

Finally, the $C(N_{f},N_{b})$ of $\alpha$-clustered $^{16}$O + $^{16}$O is compared with the results of Woods-Saxon type one, which shows visible difference between the two configurations. In the viewpoint of experimental measurements, we compare $^{16}$O + $^{16}$O results with $^{14}$N + $^{14}$N and $^{19}$F + $^{19}$F ones. It is found the $C(N_{f},N_{b})$ of $\alpha$-clustered $^{16}$O + $^{16}$O is smaller than the Woods-Saxon configured $^{16}$O + $^{16}$O. It can be understood due to the smaller relative fluctuation of the tetrahedron-configured $^{16}$O + $^{16}$O collision. Taken the initial participant nucleon into consideration, it seems the tetrahedral configuration $^{16}$O is more uniform and symmetrical comparing to the Woods-Saxon one in the AMPT model. In one word, 
the FB correlation $C(N_{f},N_{b})$ could be proposed as a probe to distinguish the exotic $\alpha$-clustering pattern experimentally and detailed measurement of pseudorapidity correlation in light nucleus collision also provides new constraints on the longitudinal dynamics of multiple parton interaction processes in the models.
In light of the present study, we look forward to the related experiments being carried out in the future LHC experiments.

\begin{acknowledgements}
	
This work was supported in part  the National Natural Science Foundation of China under contract Nos.  11890710, 11890714, 11875066, 11925502, 11961141003, the Strategic Priority Research Program of CAS under Grant No. XDB34000000, National Key R\&D Program of China under Grant No. 2016YFE0100900 and 2018YFE0104600, and by Guangdong Major Project of Basic and Applied Basic Research No. 2020B0301030008.

\end{acknowledgements}

\end{CJK*}	
\bibliography{reference}
\end{document}